\documentclass[twocolumn,showpacs,preprintnumbers,amsmath,amssymb,pre]{revtex4}
\usepackage{times}
\usepackage[sort&compress]{natbib}
\usepackage[dvips]{graphicx}

\begin{document}

\title{Exact results in the large system size limit for the dynamics of the
Chemical Master Equation, a one dimensional chain of equations}
\author{A. Martirosyan$^{1}$}
\author{David B. Saakian$^{2,3,4}$}
\email{saakian@yerphi.am}
\affiliation{$^1$ Yerevan State University, Alex Manoogian 1,
Yerevan 375025, Armenia } \affiliation{$^2$Institute of Physics,
Academia Sinica, Nankang, Taipei 11529, Taiwan}
\affiliation{$^3$Yerevan Physics Institute, Alikhanian Brothers St.
2, Yerevan 375036, Armenia} \affiliation{$^4$National Center for
Theoretical Sciences:Physics Division, National Taiwan University,
Taipei 10617, Taiwan}

\date{\today}

\begin{abstract}
We apply the Hamilton-Jacobi equation (HJE) formalism to solve the
dynamics of the Chemical  Master Equation (CME).  We found exact
analytical expressions (in large system-size limit) for the
probability distribution, including explicit expression for the
dynamics of variance of distribution. We also give the solution
for some
 simple cases of the model with time-dependent rates.
We derived the results of  Van Kampen method from HJE approach
using a special ansatz.
  Using the Van Kampen method, we give a system of ODE to define the variance in 2-d case.
We performed numerics for the CME with stationary noise. We give
analytical criteria for the
 disappearance of bi-stability in case of stationary noise in 1-d  CME.
\end{abstract}
 \maketitle
\section{Introduction}

 The statistical physics
of a living cell requires a theory for chemical reactions with few
molecules \cite{ko08},\cite{qi08}. One of mathematical tools here
is the Chemical Master Equation (CME) \cite{ga,ka92}, describing
the dynamics of probability  $P(X,t)$ of having different
(integer) number $X$ of molecules. The same CME equation could be
applied to other areas of science as well, even in the financial
market theory \cite{lu09}.
 Recently in \cite{qi09} has been considered  a
kinetic model \cite{fe01} for the
phosphorylation-dephosphorylation cycle in the cell, and the
corresponding CME was investigated. The authors considered the
model with the bi-stability phenomenon, and have been derived some
results for the existence of bi-stability. Here we solve exactly
the dynamics of CME, a very important topics according to
\cite{qi09}. The accurate (exact) solution of the CME dynamics is
important for financial market modelling \cite{lu09}. In this
article we will derive exact dynamics for the CME.

The master equation is formulated as a system of linear
differential equations for
 $P(X,t)$of having $X$ molecules, $0\le X\le N$, where N is a
 large integer:
\begin{eqnarray}
\label{e1} \frac{d{P(X,t)}}{Ndt}=
R_+(\frac{X-1}{N})P(X-1,t)+\nonumber\\
R_-(\frac{X+1}{N})P(X+1,t)-(R_+(\frac{X}{N})+R_-(\frac{X}{N}))P(X,t)
\end{eqnarray}
Here $R_+$ is the growth rate and $R_-$ is the degradation rate.

Actually we should modify the equation at the border:
\begin{equation}
\label{e2} \frac{d{P(0,t)}}{Ndt}=
R_-(\frac{1}{N})P(1,t)-R_+(0)P(0,t)
\end{equation}
and
\begin{equation}
\label{e3} \frac{d{P(N,t})}{Ndt}=
R_+(1-\frac{1}{N})P(1-\frac{1}{N},t)-R_-(1)P(N,t)
\end{equation}
The large parameter N describes the system volume.

 A close master
equations have been considered in evolution theory in
\cite{hu88}-\cite{sa08}.

Let us introduce the coordinate $x$ and function $p(x,t)$:
\begin{equation}
\label{e4} x=X/N,\quad p(x,t;N)\equiv NP(X,t)
\end{equation}
Assuming that the probability distribution is a smooth function of
$x$,
\begin{equation} \label{e5}  P(X+1,t)-P(X\pm 1,t)\ll 1
\end{equation}
one gets the Fokker-Plank equation for the model by Eq.(1).

The investigation of CME via Fokker-Plank equations meet some
problems \cite{ke}. An alternative approach is to assume that
$p(x,t)$ is not a smooth function of $x$, but the function
$u(x,t)$ is, where
\begin{equation}
\label{e6}  p(x,t;N)=\exp[Nu(x,t)]
\end{equation}
Thus the $p(x,t;N)$ might be un-smooth in the limit of
$N\to\infty$, but still have smooth u(x,t).
 Such an ansatz has
been assumed for the statics in \cite{hu88}, and for the dynamics
in \cite{sa07,ka07}, while considering the evolution models. This
ansatz gives the solution of the dynamics and steady state with
the accuracy $O(1/N)$, while the approximation of the master
equation via  Fokker-Plank equation, assuming the smoothness of
$p(x,t)$, is certainly wrong. This topic we already discussed in
\cite{sa08}, then in \cite{as10}.

In section II we calculate the dynamics of variance for population
distribution, using HJE. In section III we solve the CME dynamics
for the simple case of time dependent rates.
 In appendix we re-derive the variance of
distribution using Van Kampen method, also used that method to
calculate the variance for general 1-d multi-step CME, as well as
give ODE to derive the variance in case of 2-d CME.

\section{The master equations with constant rates}

\subsection{Hamilton-Jacobi equation for the Chemical Master
Equation} Using ansatz by Eq. (6), the model equations (1) can be
written as Hamilton-Jacobi equations for $u\equiv \ln p(x,t)/N$
\begin{equation}
\label{e7} \frac{\partial u}{\partial t}+H(u',x)=0,
\end{equation}
where $u'= \partial u/\partial x$,
\begin{eqnarray}
\label{e8} H(u',x)=R_+(x)+R_-(x)-R_+(x)e^{-u'}-R_-(x)e^{u'}
\end{eqnarray}
Eqs. (7),(8) has been derived in \cite{es09},\cite{as10}, where has
been investigated mainly the corresponding Hamilton equations to
calculate the time of extinction from meta-stabile state. We are
interested in the investigation of the whole distributions, using
the traditional mathematical method of characteristics. To solve the
HJE, we consider the Hamilton equation for x and corresponding
momentum, getting the system for the characteristics
\cite{me98,ev02}
\begin{eqnarray}
 \label{e9}
\dot x =H_v(x,v)=R_+(x)e^{-v}-R_-(x)e^{v},\nonumber\\
\dot v = -H_x(x,v),\nonumber \\
\dot u = v\,H_v(x,v)-H(x,v)=v\dot x+q,
\end{eqnarray}
subject to initial conditions: $x(0)=x_0$, $v(0)=v_0(x_0),$ $u(0) =
u_0(x_0).$ Here $v:=\partial u/\partial x,$ $ v_0(x):= u_0'(x),$
$q:=\partial u/\partial t.$

The respective solution to Eq. (\ref{e9}) in $(x,t)$ - space is
called the {\it characteristic} of Eq. (\ref{e7}).

Our Hamiltonian is time independent. Then Eq. (\ref{e7}) and Eq.
(\ref{e9}) result in
\begin{equation}
\label{e10} \dot q=0
\end{equation}
 Along the characteristic $x=x(t)$ the variable $q$ is
constant, so $q$ is selected to parameterize these curves.

Consider the equation
\begin{equation}
\label{e11}q=-(R_+(x)+R_-(x))+R_+(x)e^{-v}+R_-(x)e^{v}
\end{equation}
It has a solution for
\begin{eqnarray}
\label{e12} q\ge 0,
\end{eqnarray}
if  at some  point
\begin{equation}
\label{e13}R_+(x)=R_-(x),
\end{equation}
and we take $q\ge 0$.

Using Eq. (11),
 we transform the first equation in (\ref{e9})
into
\begin{equation}\label{e14}
\dot x = \pm \sqrt{(q+R_++R_-)^2-4R_+R_-}
\end{equation}

 Consider the following
initial distribution
\begin{equation}\label{e15}
u_0(x) = -a(x-x_0)^2.
\end{equation}
with large $a$.

The maximum of the distribution corresponds to the point $u'=0$,
therefore $q=0$.

Thus for the maximum of distribution we should consider a
characteristic with $q=0$.

Integrating the Eq.(14), we derive:
\begin{equation}\label{e16}
T=\int_{x_0}^x\frac{dy}{R_+(y)-R_-(y)}
\end{equation}
Such equation was derived in \cite{qi09}.

We can define the dynamics of the full distribution.

 Let us define the function:
\begin{equation}\label{e17}
T(x,q)=\int_{x_0}^x\frac{dy}{\sqrt{(q+R_+(y)+R_-(y))^2-4R_+(y)R_-(y)}}
\end{equation}
To calculate $u(x,t)$ we first calculate q for the given x from
the equation
\begin{equation}\label{e18}
T(x,q)=t
\end{equation}
Eq.(18) defines an implicit function
\begin{equation}\label{e19}
q=Q(x,t)
\end{equation}

\subsection{The elasticity} It is important to calculate $u''(x,t)$
at the point of the maximum of distribution, the "elasticity"
\cite{qi09}.

We  calculate $q'_x\equiv
\partial^2 u/\partial x\partial t$ from Eq.(19):
\begin{equation}\label{e20}
\frac{\partial q}{\partial x}=\frac{\partial Q(x,t)}{\partial x}
\end{equation}
We can calculate the last derivative at fixed t from the
expression:
\begin{eqnarray}\label{e21}
T(x,q)=const\nonumber\\
\frac{\partial T(x,q)}{\partial x}+\frac{\partial T(x,q)}{\partial
q} q'_x=0
\end{eqnarray}

It is equivalent to calculate $q'_x$ from Eq. (17) at the fixed t.

Thus we get the following equation for $q'$ at the point of
maximum ($u'(x,t)=0$):
\begin{eqnarray}\label{e22}
\frac{1}{R_+(x)-R_-(x)}-q'\int_{x_0}^x\frac{dy(R_+(y)+R_-(y))}{[(R_+(y)-R_-(y))]^3}=0
\end{eqnarray}
From Eq.(11) we can obtain:
\begin{eqnarray}
 \label{e23}
q'_x =-(R_+(x)-R_-(x))v'
\end{eqnarray}
Thus eventually we get:
\begin{eqnarray}\label{e24}
\frac{-1}{v'_x}=(R_+(x)-R_-(x))^2\int_{x_0}^x\frac{dy(R_+(y)+R_-(y))}{[(R_+(y)-R_-(y))]^3}
\end{eqnarray}
Eq. (24) is the main result of our work.

In Fig. 1 is given the comparison of analytical result with the
numerics.

\subsection{Probability distribution}

Eq.(17) defines the $q(x,t)$ for given $x,t$, we can define
$v(x,t)$ also using Eq.(7).

To calculate $u(x,t)$ at the given $(x,t)$, let us consider the
trajectory of points $x(\tau),\tau)$ connecting that point with
the starting point $(x_0,0)$.

We have chosen the trajectory to have $q(x(\tau),\tau)=q$. We take
$x(\tau)$ just as a solution of the equation
\begin{eqnarray}\label{e25}
\tau =T(x(\tau),q)
\end{eqnarray}
At any point of our trajectory $x(\tau),\tau)$, we can calculate
$v(x(\tau),\tau)$, while $q(x(\tau),\tau)=q$ is constant. Eq.(11)
gives
\begin{eqnarray}\label{e26}
R_-e^{2v}-(q+R_++R_-)e^v+R_+=0,\nonumber\\
v(y)=-\ln 2R_-+\ln[((q+R_+(y)+R_-(y)\pm\nonumber\\
\sqrt{(q+R_+(y)+R_-(y))^2-4R_+(y)R_-(y))}]
\end{eqnarray}
where we denoted $x(\tau)=y$.

We derive the solution of the original Eq.  (\ref{e2}),
integrating the equation $\dot u = v\dot x+q$ along our trajectory
(the characteristics connecting the points $(x,t)$ and $(x_0,0)$):
\begin{eqnarray}\label{e27}
u(x,t)=u(x,0)+\int_{x_0}^xdyv(y)+qt=\nonumber\\
\int_{x_0}^xdy[-\ln 2R_-(y)+\ln[((q+R_+(y)+R_-(y)\pm\nonumber\\
\sqrt{(q+R_+(y)+R_-(y))^2-4R_+(y)R_-(y))}]] +qt
\end{eqnarray}
Having the expression $u(x,t)$ we can calculate $p(x,t)$.

\subsection{The restricted meaning of probability distributions in
master equation}

In case of evolution models [9-13], we have master equation
similar to Eqs.(1)-(3), only the negative term $\sim P(X,t)$ has
other coefficient than $-(R_++R_-)$, and therefore there is no a
balance condition

 Contrary to the case of master
equations in evolution models \cite{hu88}-\cite{sa08} where all
the initial distributions have a meaning, now there are some
restrictions.
 We should clarify well
the meaning of the probabilities $P(X,t)$. At every moments of
time the system has only ONE value of X, and $P(X,t)$ just gives
such probabilities. We should solve the system (1),(2),(3) for the
given initial value $P(X_0,t)=1$ and $P(X,t)=0$ for other $X$,
other initial distributions have not meaning.

Another difference is connected with the stable point solutions.
Eq. (13) gives the steady state solution. If that equation has two
stable solutions $x_1,x_2$  and the probability of these two
positions is the same, i.e.
\begin{eqnarray}
\label{e28} \int_{x_1}^{x_2}dx\log\frac{R_+(x)}{R_-(x)}=0
\end{eqnarray}
we again should accurately interpret the HJE results \cite{qi09}.
For the rates given by Eq.(27) at  $a=0,k_2=10$, the transition is
at $k_1=43.1274$. One can use the HJE method to calculate the mean
period of time that solution, initially located at one stationary
point, will move to the other stationary point \cite{qi09a}. Then
it again should return back, as every moment the system could
exists only with one value of X.

In case of evolution model, the system goes to the equilibrium
state instead of oscillation between two stable solutions.

\subsection{The dynamics for the stationary but random rates.}
Consider now the case when the rates are smooth functions of $x$
plus some random noises. The noise in the rates is well confirmed
experimentally \cite{xi10}.
 We took the case of rates from
\cite{qi09},
\begin{eqnarray}\label{e29}
R_+(k_1,x)=(1 - x) (0.5+ k_1 x^2),\nonumber\\
R_-(k_2,x)=x (k_2 + 0.01 x^2)
\end{eqnarray}
where now $k_19$ and $k_2$ are random variables,
\begin{eqnarray}\label{e30}
k_1= K_1\exp(a\xi_1(x)-a^2/2) x^2),\nonumber\\
k_2=K_2\exp(a\xi_2(x)-a^2/2)
\end{eqnarray}
$\xi_1(x),\xi_2(x)$ have normal distributions. We consider the
model with $N=100$, and performed numerics for different values of
parameters $K_1,K_2,a$. For $K_1=50,K_2=10$, at $a\approx 0.9$
there is a phase transition: instead of two steady state solution
we get one steady state $x\approx 0.088$.

We can analytically estimate the transition point (from one
stabile point to bi-stability), considering the behavior of
\begin{eqnarray}\label{e31}
U(x)=<\ln\frac{R_+(k_1,x)}{R_-(k_2,x}>|_{k_1,k_2}
\end{eqnarray}
We found that the function $U(x)$ changes its behavior with the
level of the noise. At $a=0$ it has only three roots $U(x)=0$,
while for $a> 0.75$ there is one root.

\begin{figure} \large \unitlength=0.1in
\begin{picture}(42,12)
\put(0,0){\includegraphics{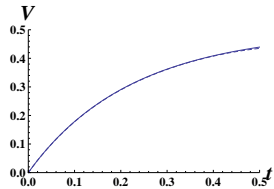}}
\end{picture}
\caption{The graphics for the elasticity $V(t)\equiv
\frac{-1}{v'(t)}$ for the model with
$N=100,R_+(x)=\exp(-x),R_+(x)=\exp(x),x(0)=0.5$. The smooth line
is the analytical result by Eq.(24) and the dashed line is the
numerical result. The difference is less than $0.5$\%. }
\end{figure}

\begin{figure} \large \unitlength=0.1in
\begin{picture}(42,12)
\put(0,0){\includegraphics{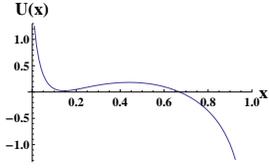}}
\end{picture}
\caption{The graphics for the $U(x)=\ln (R_+(x)/R_-(x))$  at
$K_1=44,K_2=10,a=0$. There are three solutions for the equation
$U(x)=0$. }
\end{figure}

\begin{figure} \large \unitlength=0.1in
\begin{picture}(42,12)
\put(0,0){\includegraphics{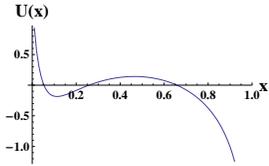}}
\end{picture}
\caption{The graphics for the $U(x)=<\ln (R_+(x)/R_-(x))>$ at
$K_1=44,K_2=10,a=0.8$. There is only one solution for the equation
$U(x)=0$. }
\end{figure}

\section{Master equation when the rates vary with time}
\subsection{The simplest solvable case}

Consider the case when birth and death rate coefficients in CME
change with the time as $g(t)$ and $f(t)$, while they are the same
for all the $x$. We have the following Hamiltonian:
\begin{equation}\label{e32}
-H = g(t) e^{-v} + f(t)e^{v}-g(t)-f(t), \quad v = \frac{\partial
u}{\partial x}
\end{equation}
We get for the characteristics the following system of equations:
\begin{equation}\label{e33}
\frac{dp}{dt} = - \frac{\partial H}{\partial x} = 0
\end{equation}
Thus the $p$ is constants:
\begin{equation}\label{e34}
v=v_0,
\end{equation}
For the coordinate we have a simple equation:
\begin{equation}\label{e35}
\frac{dx}{dt} = \frac{\partial H}{\partial v} = g(t) e^{-v_0} -
f(t) e^{v_0}
\end{equation}
Thus we have a simple solution
\begin{equation}\label{e36}
x-x_0 =
 e^{-v_0}\int_0 ^{t} g(\tau) - e^{v_0} \int_0^{t} f(\tau)
\end{equation}
We find $v_0$ as a function of $x,t$ from the solution of the last
equation.
\begin{eqnarray}\label{e37}
u(x,t) = u(x_0,0)+v_0 (x-x_0) +\nonumber\\
 \int_0^t [ g(\tau)
e^{-v_0} + f(\tau)e^{v_0}-g(\tau)-f(\tau) ] d\tau
\end{eqnarray}
For the maximum of distribution we have $v_0=0$, therefore we get
an equation:
\begin{equation}\label{e38}
x-x_0 =
 \int_0 ^{t}[ g(\tau) -  f(\tau)]d\tau
\end{equation}
It is interesting to find the variance of distribution.
Differentiating Eq. (36), we get, putting $v_0=0$ at the maximum
point:
\begin{equation}\label{e39}
-v'(x,t)= \frac{1}{\int_0 ^{t} [g(\tau) +f(\tau)]}d\tau
\end{equation}
Thus the variance of distribution decreases with the time.

\subsection{Rates as linear functions of the number of molecules}

Consider now the Hamiltonian
\begin{equation}\label{e40}
-H = (ax+ g(t)) e^{-v} + (bx + f(t))e^{v}-(a+b)x-f(t)-g(t)
\end{equation}
We have
\begin{equation}\label{e41}
\frac{dv}{dt} = a e^{-v} + b e^{v} - (a+b)
\end{equation}
We can solve this case:
\begin{eqnarray}\label{e42}
\int_{v_0}^v \frac{dv}{a e^{-v} + b e^{v} - (a+b)}=t\nonumber\\
\end{eqnarray}
We can define the function $v=V(v_0,t)$ from the latter equation
\begin{eqnarray}
\label{e43} V(v_0,t)=\text {Log}\left[\frac {a e^{a t} - a e^{b t}
- b e^{a t + \text {v0}} + a e^{b t + \text {v0}}} {a e^{a t} - b
e^{b t} - b e^{a t + \text {v0}} + b e^{b t + \text {v0}}} \right]
\end{eqnarray}
We have also
\begin{eqnarray}
\label{e44} \frac{dV}{dv_0}(t)= \frac {-b e^{a t + v_0} + a e^{b t
+ v_0}} {a e^{a t} - a e^{b t} -
    b e^{a t + v_0} + a e^{b t + v_0}} -\nonumber\\
     \frac {-b e^{a t + v_0} +
    b e^{b t + v_0}} {a e^{a t} - b e^{b t} - b e^{a t + v_0} +
    b e^{b t + v_0}}
\end{eqnarray}
 Now we solve the equation for x:
\begin{equation}\label{e45}
\frac{dx}{dt} =  (b-a)x + f(t)e^{V}- e^{-V}g(t)
\end{equation}
Its solution gives
\begin{equation}\label{e46}
x-x_0 = e^{(b-a)t} \int_0^t[ f(\tau)e^{V(v_0,\tau)}-
g(\tau)e^{-V(v_0,\tau)}]e^{-(b-a)\tau}d\tau
\end{equation}
Eqs.(42),(46) together define the trajectory of characteristics.

To get the solution for the maximum, we put $v=0$ on the left hand
side of Eq. (42). In this way we get a simple explicit equation
for $v_0$ as a function of time t.
\begin{eqnarray}
\label{e47}\frac {a e^{a t} - a e^{b t} - b e^{a t + \text {v0}} +
a e^{b t + \text {v0}}} {a e^{a t} - b e^{b t} - b e^{a t + \text
{v0}} + b e^{b t + \text {v0}}}=1
\end{eqnarray}
Putting that solution into Eq.(46), we find the  trajectory of the
maximum of distribution.

To calculate $v'$ we should differentiate the expression in
Eq.(46) via x. Using the relation $V'(v_0,t)=1$, we derive
\begin{eqnarray}\label{e48}
\frac{1}{v'} =e^{(b-a)t}\times
\nonumber\\
\int_0^t[ f(\tau)e^{V(v_0,\tau)}+
g(\tau)e^{-V(v_0,\tau)}]\frac{dV}{dv_0}(\tau)e^{-(b-a)\tau}d\tau
\end{eqnarray}
where $v_0$ is defined by Eq.(46) as a function of t.

\section{CME in multi-dimensional space}

 Consider now the CME in multi-dimensional space,
when we have d-dimensional $\vec X$.
\begin{eqnarray}
\label{e49} \frac{d{P(\vec X,t)}}{Ndt}=\nonumber\\
\sum_{n_l=-K}^{K}R_{\vec n}(\frac{\vec X-\vec n}{N})P(\vec X-\vec
n,t) -\sum_{\vec n}R_{\vec n}(\frac{\vec X}{N})P(\vec X,t)
\end{eqnarray}
We get HJE for the function $u(\vec \vec x)$ with following
Hamiltonian:
\begin{eqnarray}
\label{e50} \frac{\partial H}{\partial t}+H(\vec x,\vec
p)=0,\nonumber\\
- H(\vec x,\vec p)=\sum_{n_l=-K,1\le l\le d}^{K}R_{\vec n}(\vec x
)[\exp[-\vec n \vec p]-1]
\end{eqnarray}
Let us investigate the motion of the maximum of distribution.
 We denote $\vec y(t)=\vec x $ the maximum of distribution at
 moment of time t, and assume
the following ansatz for the $u(x,t)$ near the maximum of
distribution:
\begin{eqnarray}
\label{e51} u(\vec x,t)=-\frac{1}{2}&lt;\vec x-y(t)|V|x-y(t)&gt;
\end{eqnarray}
Putting this ansatz into
\begin{eqnarray}
\label{e52} u''_{x_lt}+H'_{x_l}(\vec x,\vec p)+\sum_m H'_{\vec
p_m}(\vec x,\vec p)\frac{d p_m}{d x_l}=0
\end{eqnarray}
Consider the point $\vec x=\vec y(t),\vec p=0$, and use the
identity:
\begin{eqnarray}
\label{e53} \frac{d p_m}{d x_l}=-V_{ml},
\end{eqnarray}
We get:
\begin{eqnarray}
\label{e54} \sum_mV_{lm}\frac{dy_l}{dt}
-(\sum_{n_h=-K}^KR_{n_1...n_d }\sum_mn_m)V_{lm}=0
\end{eqnarray}
 We get the following ODE for the dynamics of the maximum of
 distribution:

\begin{eqnarray}
\label{e55} \frac{dy_l(t)}{dt}=\sum_mQ_{m} y_m(t)),\nonumber\\
Q_{m}=\sum_{1\le h\le d }\sum_{-K\le n_h\le K}R_{n_1..n_d}n_m
\end{eqnarray}

\section{Derivation of elasticity in 1-d case}

Let us get system of ODE for the $V$. We consider the multi-step
version of CME in 1-d.

\begin{eqnarray}
\label{e56}
 \frac{dP(X,t)}{Ndt}=
\nonumber\\
\sum_{n=-K}^{K}R_{n}(\frac{X-n}{N})P(X-n,t) -
\sum_{n}R_{n}(\frac{X}{N})P( X,t)
\end{eqnarray}

Now we have the following Hamilton-Jacobi equation:
\begin{eqnarray}
\label{e57}- H(u',x)=\sum_{n=-K}^{K}R_{n}(x)[e^{-nu'} -1]
\end{eqnarray}
Now we have to consider the higher terms in expansion of $u(x,t)$
near the maximum.
\begin{eqnarray}
\label{e58}
 u=-V(x-y(t))^2/2-T(x-y(t))^3/6
\end{eqnarray}
Eq.(55) gives
\begin{eqnarray}
\label{e59} \frac{dy(t)}{dt}=y(t))b\nonumber\\
b=\sum_{-K\le n\le K}R_{n}(x)n
\end{eqnarray}
We also denote:
\begin{eqnarray}
\label{e60} a(x)=\sum_{-K\le n\le K}R_{n}(x)n^2
\end{eqnarray}
therefore
\begin{eqnarray}
\label{e61} \dot{y}=b\nonumber
\end{eqnarray}
 With the ansatz by Eq.(56) we have:
\begin{eqnarray}
\label{e62} u'''_{xxt}=-\dot{V}+Tb,\nonumber\\
u'''_{xtt}=2\dot{V}b-Tb^2+V\dot{\dot{y}}=\nonumber\\2\dot{V}b-Tb^2
+Vbb'
\end{eqnarray}
From the other hand we have differentiating the right hand side of
Eq.(50):
\begin{eqnarray}
\label{e63}-u'''_{xxt}=H''_{pp}V^2-2H''_{xp}V-TH'_p=aV^2-2b'V-bT,\nonumber\\
-u'''_{xtt}=
-H''_{xp}V\dot{y}+H''_{pp}bV^2-H'_{p}\dot{V}+H'_pTb^2=\nonumber\\
-bb'V+abV^2-b\dot{V}+Tb^2
\end{eqnarray}
 We derived Eqs.(58),(59) putting $x=y(t)$.

Then
\begin{eqnarray}
\label{e64} -\dot{V}+2Tb=aV^2-2b'V,\nonumber\\
3\dot{V}b-2Tb^2=aV^2b-2b'V\nonumber
\end{eqnarray}
Removing T we get
\begin{eqnarray}
\label{e65} 2b\frac{d}{dt}V=-4bb'V+2baV^2
\end{eqnarray}
or
\begin{eqnarray}
\label{e66} \frac{d}{dt}\frac{1}{V}=b'\frac{1}{V}-2a
\end{eqnarray}
Which gives Eq.(24) for 1-step CMW case.

For the multi-step CME we have the following expression for the
elasticity:
\begin{eqnarray}\label{e67}
\frac{-1}{v'_x}=(b(x))^2\int_{x_0}^x\frac{dya(y)}{[b(y)]^3}
\end{eqnarray}

\section{Conclusion}
In conclusion, we calculated exact probability dynamics Eq.(27)
for the Chemical Master Distribution using Hamilton Jacobi
equation method. Using the methods of characteristcs for solution
of HJE, we gave explicit expression for the variance of
distribution Eq.(24). The latter is important both for chemical
\cite{qi08} and financial applications. We derived the variance
also directly from HJE, using a special ansatz. The latter is
equivalent to Van Kampen method \cite{ka92}. Using Van Kampen
method, in Appendix we give an exact expression for the variance
for general 1-d model, as well as a system of ODE to define the
variance in 2-d case.  Both HJE and Van Kampen methods gave
identical results for the variance and the maximum point dynamics
of CME: HJE gives directly differential equation for the vasrianc,
while the Van Kampen method originally gives a differential
equation for the probability distributions, Eq.(A.5), which later
proved the differential equation for the variance. HJE gives also
exact steady state distribution, and is more adequate for
investigation of meta-stabile points \cite{es09} \cite{as10}.
Using HJE, we derived exact dynamics Eqs.(44)-(46) for the case of
1-d CME when the rates are linear functions of the coordinate
(number of molecules) plus some time dependent functions.

We performed some numerics in case of static noise, also give an
analytical criteria for the level of the noise when the
bi-stability disappears. Our choice of potential for the averaging
Eq.(31) is rather arbitrary (contrary to all other results in this
article, which are derived rigorously and are exact). A further
investigation of the problem is necessary. Perhaps it is possible
to investigate the more realistic case of non-stationary noise
\cite{xi10}.

DBS thanks DARPA Prophecy Program and Academia Sinica for
financial support.

 \renewcommand{\theequation}{A.\arabic{equation}}
\setcounter{equation}{0}
\appendix

\section{The calculation of variance dynamics using the Van Kampen method}
\subsection{1-d multi-step models.}

Consider the CME defined trough the system of equations:
\begin{eqnarray}
\label{a1} \frac{d{P(X,t)}}{Ndt}=\nonumber\\
\sum_{\l=-K}^{K}R_l(\frac{X+l}{N})P(X+l,t)
-\sum_lR_l(\frac{X}{N})P(X,t)
\end{eqnarray}
Following to Van Kampen, we assume that near the maximum
\begin{eqnarray}
\label{a2} X=N\phi(t)+\sqrt{N}\xi(t),\nonumber\\
P(X,t)=\Pi(\xi,t)
\end{eqnarray}
 For the maximum point
$\phi(t)=\sum_iiP(i,t)$ we will derive an equation (4) later.

 We consider terms with different
scaling by $N$ in the equation:
\begin{eqnarray}
\label{a3} \frac{d{\Pi(\xi,t)}}{dt}-\sqrt{N}\phi'(t)\frac{\partial \Pi}{\partial \xi}=\nonumber\\
N\sum_{\l=-K}^{K}R_l(\phi(t)+\frac{\xi}{\sqrt{N}}-\frac{1}{N})P(X-l,t)\nonumber\\
-N\sum_lR_l(\phi(t)+\frac{\xi}{\sqrt{N}})P(X,t)
\end{eqnarray}
Collecting together the $\sim \sqrt{N}$ terms, we get
\begin{eqnarray}
\label{a4} \frac{d\phi(t)}{dt}= \sum_{l=-K}^{K}lR_l(\phi(t))
\end{eqnarray}
The $N^0$ terms give an equation
\begin{eqnarray}
\label{a5} \frac{d\Pi(t)}{dt}=-b'\frac{\partial }{\partial \xi}(
\xi)\Pi+
 \frac{a}{2}
 \frac{\partial^2 }{\partial
\xi^2}\Pi\nonumber\\
b=\sum_llR_l,a=\sum_ll^2R_l
\end{eqnarray}
We derive the following equations for $<\xi_i>$:
\begin{eqnarray}
\label{a6} \frac{d<\xi>}{dt}=b'(t)<\xi_i>
\end{eqnarray}
The initial condition $<\xi(0)>=0$ gives $<\xi(t)>=0$.

We get for the variance
\begin{eqnarray}
\label{a7}
 \frac{d}{dt}<\xi(t)^2>=
 a(\psi(t))
 +2b'(\psi(t))
 <\xi_i(t)>)^2
\end{eqnarray}
 We can consider ODE
via $\psi$ instead of t. Then Eq. (A.4) gives $d\psi/dt=b$, then:
\begin{eqnarray}
\label{a8}
 b\frac{d}{d\psi}<\xi^2>=
 a(\psi)+2b'(\psi)
 <\xi>)^2
\end{eqnarray}
Eventually we get Eq.(24), with $R_+-R_-\to b\equiv
\sum_lR_ll,R_++R_-\to a\equiv\sum_ll^2R_l$.

\subsection{2-d case}

We will apply the Van Kampen method, giving  ODE to calculate the
variance of distribution in 2-d case.

Consider the following system of equations for $P(X,Y,t),0\le X\le
N,0\le Y\le N$:
\begin{eqnarray}
\label{a9} \frac{dP(X,Y,t)}{dt}=
-(R_{1+}(\frac{X}{N},\frac{Y}{N})+R_{1-}(\frac{X}{N}),\frac{X}{N})\nonumber\\
+R_{2+}(\frac{X}{N},\frac{Y}{N})+R_{2-}(\frac{X}{N}),\frac{X}{N}))P(X,Y,t)+\nonumber\\
R_{1+}(\frac{X-1}{N},\frac{Y}{N})P(X-1,Y,t)+\nonumber\\
R_{1-}(\frac{X+1}{N},\frac{Y-1}{N})P(X+1,Y,t)\nonumber\\
R_{2+}(\frac{X}{N},\frac{Y-1}{N})P(X,Y-1,t)+\nonumber\\
R_{2-}(\frac{X}{N},\frac{Y+1}{N})P(X,Y+1,t)
\end{eqnarray}
Following to \cite{ka92}, we introduce the fluctuating variables
$\xi_1,\xi_2$ and replace $P(X,Y,t)$ by $\Pi(\xi_1,\xi_2,t)$,see
\cite{ka92}:
\begin{eqnarray}
\label{a10} X=N\phi_1(t)+\sqrt{N}\xi_1(t),\nonumber\\
Y=N\psi_2(t)+\sqrt{N}\xi_2(t),\nonumber\\
P(X,Y,t)=\Pi(\xi_1,\xi_2,t)
\end{eqnarray}
where $\psi_1(t), \psi_2(t)$ give the solution in case of infinite
$N$:
\begin{eqnarray}
\label{a11} \frac{d\psi_1(t)}{dt}=b_1(\psi_1(t),\psi_2(t))\nonumber\\
\frac{d\psi_2(t)}{dt}=b_2(\psi_1(t),\psi_2(t))\nonumber\\
b_1(\psi_1,\psi_2)=R_{1+}(\psi_1,\psi_2)-R_{1-}(\psi_1,\psi_2),\nonumber\\
b_2(\psi_1,\psi_2)=R_{2+}(\psi_1,\psi_2)-R_{2-}(\psi_1,\psi_2)
\end{eqnarray}
We can solve Eq.(A.11) and calculate $\psi_1(t),\psi_2(t)$.

 Following to the methods of \cite{ka92}, we
derive an  equation:
\begin{eqnarray}
\label{a12}
\frac{d\Pi(t)}{dt}=\nonumber\\
-\sum_{\alpha}(R'_{\alpha+}-R'_{\alpha-})
\frac{\partial}{\partial \xi_{\alpha}}(\xi_{\alpha}\Pi(\xi_1,\xi_2,t) )+\nonumber\\
\sum_{\alpha}\frac{R_{\alpha +}+R_{\alpha
-}}{2}\frac{\partial^2}{\partial^2 \xi_{\alpha}}\Pi
(\xi_1,\xi_2,t)\nonumber\\
\equiv -\sum_{\alpha}\frac{\partial b_{\alpha}}{\partial
\xi_\alpha}\frac{\partial }{\partial
\xi_\alpha}\Pi+\frac{1}{2}\sum_{\alpha}a_{\alpha}\frac{\partial^2
}{\partial \xi_\alpha^2}\Pi
\end{eqnarray}
where we denoted $R'_{\alpha \pm }\equiv \frac{\partial R_{\alpha
\pm}(\psi_1(t),\psi_2(t))}{\partial \xi_{\alpha}},R_{\alpha
\pm}\equiv R_{\alpha \pm}(\psi_1(t),\psi_2(t)) $.

We derive the following equations for $<\xi_i>$:
\begin{eqnarray}
\label{a13} \frac{d<\xi_i>}{dt}=b'_i(t)<\xi_i>
\end{eqnarray}
The initial condition $<\xi(0)>=0$ gives $<\xi(t)>=0$.

We get for the variance
\begin{eqnarray}
\label{a14}
 \frac{d}{dt}<\xi_i(t)^2>=\nonumber\\
 a_i(\psi_1(t),\psi_2(t))
 +2b_i'(\psi_1(t),\psi_2(t))
 <\xi_i(t)>)^2
\end{eqnarray}
We can calculate the variance numerically as function of $t$,
using the solution of Eq.(A.11).

\end{document}